# Deep-learning PDEs with unlabeled data and hardwiring physics laws

S. Mohammad H. Hashemi[*,1], Demetri Psaltis[1]

Providing fast and accurate solutions to partial differential equations is a problem of continuous interest to the fields of applied mathematics and physics. With the recent advances in machine learning, the adoption learning techniques in this domain is being eagerly pursued. We build upon earlier works on linear and homogeneous PDEs, and develop convolutional deep neural networks that can accurately solve nonlinear and non-homogeneous equations without the need for labeled data. The architecture of these networks is readily accessible for scientific disciplines who deal with PDEs and know the basics of deep learning.

**Introduction**

Among various fields that have started to benefit from the boom in machine learning (ML) and artificial intelligence (AI), physics is no exceptions[1-3]. Different scenarios have been proposed for contribution of machine learning in this field. First, ML can provide independent simulation methods that are faster than the conventional technics at the price of tolerable losses in accuracy[4]. Second, it can act as a supportive tool to improve the traditional solvers in computational physics. For instance, initial solutions provided by deep neural networks (DNNs) that are very close to the final solution, can be used as a starting solution to expedite the convergence of such solvers[5]. Another example is the training of machine learning with data from expensive and highly sophisticated solvers, and then use the trained network to improve the results of simpler and lighter numerical models in an efficient way[6]. Third, deep learning (DL) can make the data collected from the experimental data richer by filling the gaps in the variables space[7].

In this paper, we focus on the application of ML in physics whereby the structure of the developed network is modified to integrate some aspects of the underlying physics of the problems[1,8]. These works are in fact building on the inherent advantage of all the problems in physics: they are governed by mathematical laws of the nature rather than the cultural or societal conventions, which are often quite difficult to formulate. These governing laws are often described in the form of partial differential equations (PDEs). As a result, providing solution for PDEs is the focus of many recent DL works[9-11]. It should be noted that the idea of using neural networks to solve PDEs is not new and there are some pioneering works[12,13] who tackled this problem even before the recent wave of ML studies.

An elegant method was recently reported, which removes the need for the labeled training data using convolutional neural networks (CNNs)[5]. In this paper, we discuss why this method - the way described in Sharma et al.[5] - only works for linear and homogeneous equations. We then explain how to generalize it for more complex PDEs by testing it first for a non-homogeneous, but linear equation and then on a non-homogeneous and nonlinear equation. Finally, the robustness of the proposed DL technique is investigated when boundary conditions (BCs) take values not presented in the training/test sets. The results of this last part indicate that our straightforward approach performs well even outside of the training variable space.

---

[*] Corresponding author: mohammad.hashemi@epfl.ch
1  Laboratory of Optics, École Polytechnique Fédérale de Lausanne (EPFL), Lausanne, Switzerland

**Hardwiring physics in kernels**

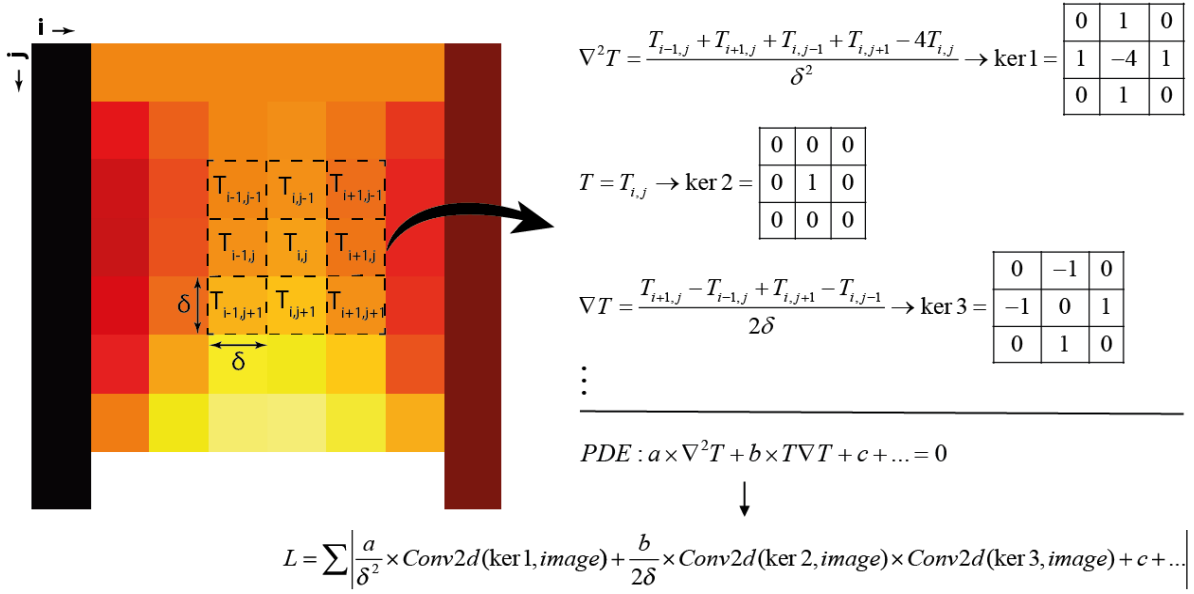

Figure 1. Convolution of physics-derived kernels with the pixels of an image that represents the solution domain of the problem's governing PDE. The form and number of differential operators in a PDE defines the loss function (L). The kernels are defined for three terms in this illustration, but the PDE can contain more terms which result in a more complicated L.

The approach of Sharma et al.[5] can be summarized as follows: a set of different Dirichlet BCs are applied to boundary pixels of images that represent the geometric domain of the governing PDE's dependent variable. The non-boundary pixels are all set to zero as initial values. These images are then fed into a DNN with U-Net structure[14] to reconstruct images at the output; each estimating the complete solution for every set of BCs found at the input. The DNN weights and biases are updated while minimizing a custom, physics-informed loss function that convolves the outputs of the network with a kernel that establishes a physical relationship between the pixels of the output pattern. This kernel's values are found through discretization of the PDE's differential operator. As an example, the authors in Ref. 5 use the Laplace equation ($\nabla^2 T = 0$) to model the heat conduction problem in a 2D square domain. They utilize only one 3 by 3 kernel - based on the finite difference discretization of the Laplacian operator - to convolve with the U-Net's output images and calculate a loss function based on the resulting convolution. The authors then claim that this method of minimizing such a loss function, i.e. convolution of the field image with a single kernel, should work for all PDEs. However, we remark that their method is not applicable to PDEs containing any of the following terms: more than one linear differential operator, a non-homogenous (source) term, and one or more non-linear operators. For such PDEs, more complex cost functions, often including multiple kernels should be defined as we illustrate in Figure 1. This figure shows how the discretized value of the dependent variable (always T in this paper) is associated with the image's pixels. In addition, the definition of few kernels for some differential operators and their contributions to the cost function of a PDE containing those terms is provided. These kernels are obtained from the finite difference discretization[15] of differential operators of various forms. Since the variable T must satisfy the PDE at all positions in the 2D output grid, the loss funtion must be evaluated at all positions. In other words, the loss function simulates the form of the PDE of interest through convoltuion between different kernels and the output pattern of the neural network. All the kernels are 3×3 in this study, but kernels with larger dimensions can be defined based on discretization for higher orders. We note that this method can be also deployed for 3D geometries using 3D grids and kernels.

**Network architecture and implementation**

The schematic structure of the U-Net DNN used in this work is illustrated in Figure 2. As the goal of this study is to provide a proof-of-concept, we chose to work with 16 by 16 pixel images to minimize computing time. For larger images, the model should minimize a linear combination of the loss function applied to the output image in addition to a set of its downsampled versions[5, 16].

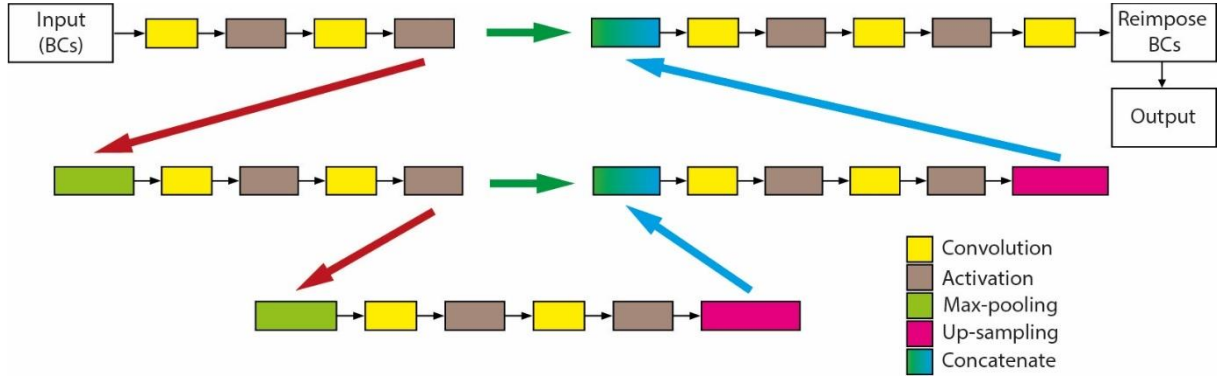

Figure 2. The architecture of the U-Net DNN employed to provide PDE solutions. The inputs to the network are some images of the solution domain with their boundary pixels equal to the BC values and all other internal pixels set to 0. The weights and biases are updated by minimizing the custom loss functions defined based on the involved physics. Before generating the output, the BCs are reapplied to the boundary pixels before launching the next forward propagation step.

We implemented the U-Net model of Figure 2 in Keras using TensorFlow backend. The activation function is Exponential Linear Unit (ELU) in all layers and Adam optimizer is used in training the model. U-Net models that are commonly used for image segmentation apply the Sigmoid activation function right before the output. However, we do not use any activation function to obtain the output. This is due to the fact that we are interested in the exact pixel values of the output as they represent the discretized values of the dependent variable, i.e. the solution.

**Results**

We test the performance of the DNN on three different equations: first on a linear and homogeneous (LH) equation, then on a linear and non-homogeneous (LNH) equation, and finally on a nonlinear and non-homogeneous equation (NLNH). 7920 images, each with a different set of four BCs in the range of [0,1] and a step size of 0.10 are generated. A typical BC set can be for instance: {L, R, B, T} = {0.20, 0.40, 0.00, 0.90}, where L, R, B, and T denote left, right, top, and bottom BC, respectively. The value of all internal pixels is set to 0. 80% of these images are fed into the DNN in the training phase and the rest are used to test its performance.

After verifying that the network can learn to solve LH, LNH, and NLNH equations with BCs of the train/test sets, its generalization performance is put into test for two cases: Case (1) where all BC values are randomly assigned within the training range, i.e. [0, 1], but without any step size constraint, and Case (2) where at least two BC values in each input image are set to be outside of the training range.

To measure the DNN's accuracy, ground truths are obtained from a high-fidelity solver based on the Finite Difference Method (FDM) that takes in the BCs as the input and solves the corresponding equation over the whole domain in an iterative manner. We emphasize that these ground truths are not used anywhere in the process of DNN training, and that they are solely used for the purpose of comparison with DNN outputs.

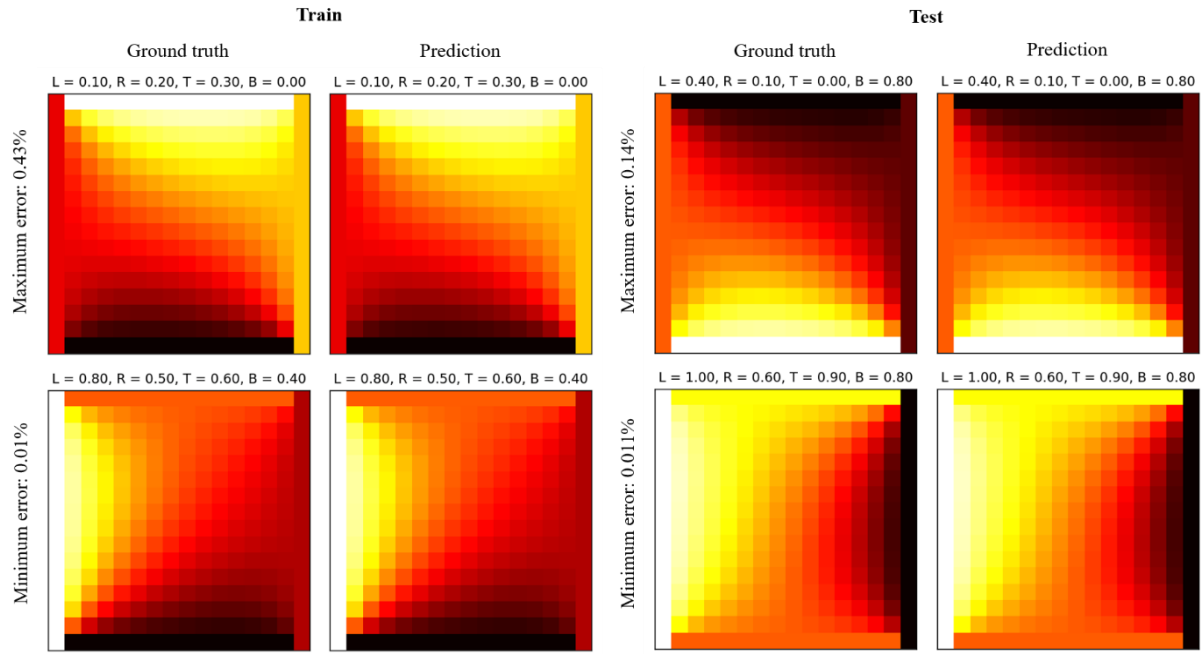

Figure 3. FDM results (ground truth) versus DNN output (prediction) for LH equation: the upper and lower rows show the results for the maximum and minimum errors within the train (left panel) and the test (right panel) BCs. The values on top of each field image shows the corresponding BC (L: Left BC, R: Right BC, T: Top BC, and B: Bottom BC).

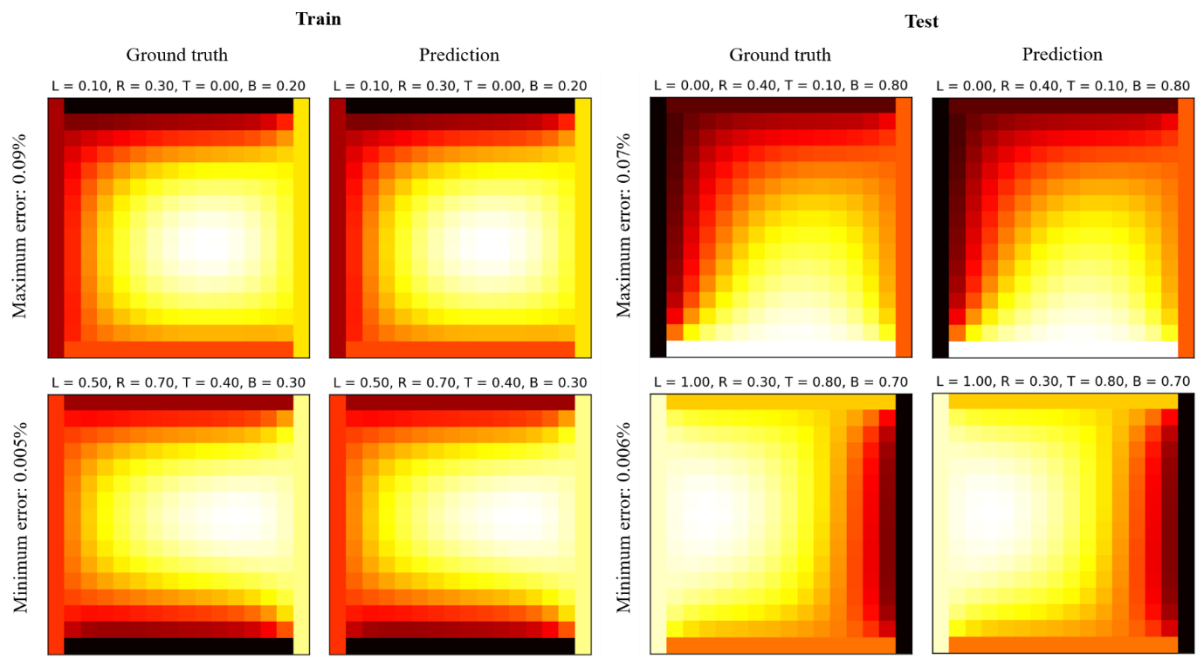

Figure 4. FDM results (ground truth) versus DNN output (prediction) for LNH equation: the upper and lower rows show the results for the maximum and minimum errors within the train (left panel) and the test (right panel) BCs. The values on top of each field image shows the corresponding BC (L: Left BC, R: Right BC, T: Top BC, and B: Bottom BC).

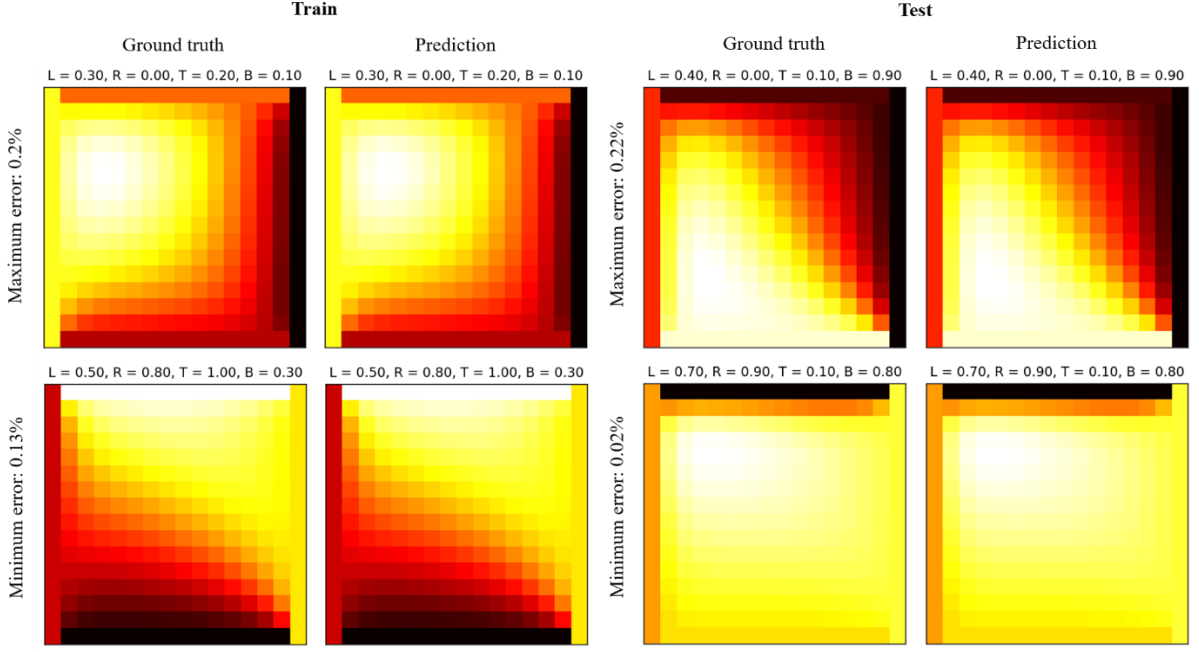

Figure 5. FDM results (ground truth) versus DNN output (prediction) for NLNH equation: the upper and lower rows show the results for the maximum and minimum errors within the train (left panel) and the test (right panel) BCs. The values on top of each field image shows the corresponding BC (L: Left BC, R: Right BC, T: Top BC, and B: Bottom BC).

*LH equation.* The homogeneous Laplacian equation ($\nabla^2 T = 0$) is considered and the loss function is defined as follows:

$$Loss = \frac{1}{\delta^2} \sum |Conv2d(\text{ker1}, output)| \quad \text{where:} \quad \text{ker1} = \begin{array}{|c|c|c|} \hline 0 & 1 & 0 \\ \hline 1 & -4 & 1 \\ \hline 0 & 1 & 0 \\ \hline \end{array}$$

In 16 by 16 pixel images $\delta = 1/16$ as both width and height of the domain is assumed to be one. The mean absolute percentage error (MAPE) of the network output - with regard to the FDM ground truths - for both the training and the test BCs is 0.04% with a standard deviation (STD) of 0.03% for the training and 0.02% for the test sets, respectively.

*LNH equation.* The ability of DNN in solving non-homogeneous equations is investigated by the Laplacian equation with a constant source term ($\nabla^2 T + 4 = 0$), resulting in the following loss function:

$$Loss = \sum \left| \frac{1}{\delta^2} Conv2d(\text{ker1}, output) + 4 \right|$$

The MAPE of the network output for both the training and the test BCs is 0.02% with equal STD of 0.01%.

*NLNH equation.* In order to verify if the DNN can handle the nonlinear PDEs as well, we add two nonlinear terms to the LNH equation: $\nabla^2 T + 16 \times T\nabla T + 4 \times T^2 + 4 = 0$. This equation does not correspond to a physical problem anymore, but is well suited for testing the technique on a nonlinear case. The loss function then takes the more complex form of:

$$Loss = \sum \left| \frac{1}{\delta^2}(Conv2d(\text{ker}1, output) + \frac{8}{\delta} \times Conv2d(\text{ker}2, output) \times Conv2d(\text{ker}3, output) + 4 \times Conv2d(\text{ker}2, output) \times Conv2d(\text{ker}2, output) + 4) \right|$$

where:  $\text{ker}2 = \begin{bmatrix} 0 & 0 & 0 \\ 0 & 1 & 0 \\ 0 & 0 & 0 \end{bmatrix}$,  $\text{ker}3 = \begin{bmatrix} 0 & -1 & 0 \\ -1 & 0 & 1 \\ 0 & 1 & 0 \end{bmatrix}$

The MAPE of the network output for the training BCs is 0.05% with STD of 0.02%. For the test set, the MAPE is 0.10% with STD of 0.04%.

Figure 3, Figure 4, and Figure 5 present the comparison between the network outputs for BCs that result in the maximum and minimum MAPEs with the corresponding ground truths for LH, LNH, and NLNH equations, respectively.

Slightly higher MAPEs for the NLNH equation results are somewhat expected, given the more complex form of its DNN's cost function that includes convolution of the output images with three kernels instead of one kernel in LH and LNH cases.

*Case 1.* So far, we only evaluated the DNNs for BC values that can be found in training or test sets, i.e. any four numbers from {0, 0.1, 0.2, …, 1} set with repetitions being authorized. In this subsection, we test the NLNH network for 300 input images whose BCs were set by random values from the range [0,1].

Calculating a MAPE of 0.07% with STD of 0.04%, we remark that the network is quite adept at generating highly accurate results for the BCs of Case 1.

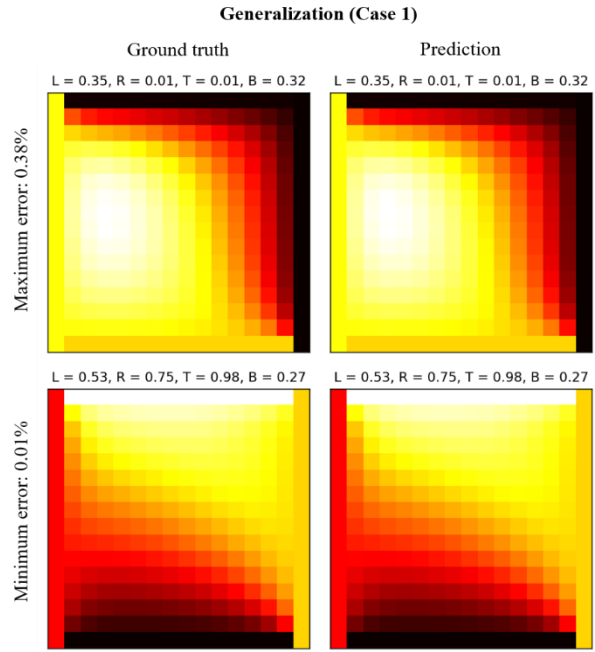

Figure 6. FDM results (ground truth) versus DNN output (prediction) for NLNH equation: the upper and lower rows show the results for the maximum and minimum errors within the 300 input images whose randomly assigned BC values are in the range of [0,1]. The values on top of each field image shows the corresponding BC (L: Left BC, R: Right BC, T: Top BC, and B: Bottom BC).

The comparison between the network outputs for BCs that result in the maximum and minimum MAPE with the corresponding ground truths is depicted in Figure 6.

*Case 2.* For a second generalization test, we generated 300 input images whose left, right, top and bottom BCs are set by random numbers from the ranges [0.4, 1.4], [0.6, 1.6], [1, 2], and [1, 2], respectively. This guarantees that in every input image, at least two out of four BCs are outside of the training range of [0, 1]. These images are then fed into the DNNs for NLNH, and LNH equations.

For NLNH equation, MAPE is 1.65% with STD of 0.99%. These values improve for LNH equation: a MAPE of 1.05% with STD of 0.51%. These values indicate very good precision of the networks when dealing with Case 2 BCs. Overall Case 1 and Case 2 results prove that integrating the physics of the problem at the hand into the DNNs, makes even their generalized predictions reliable.

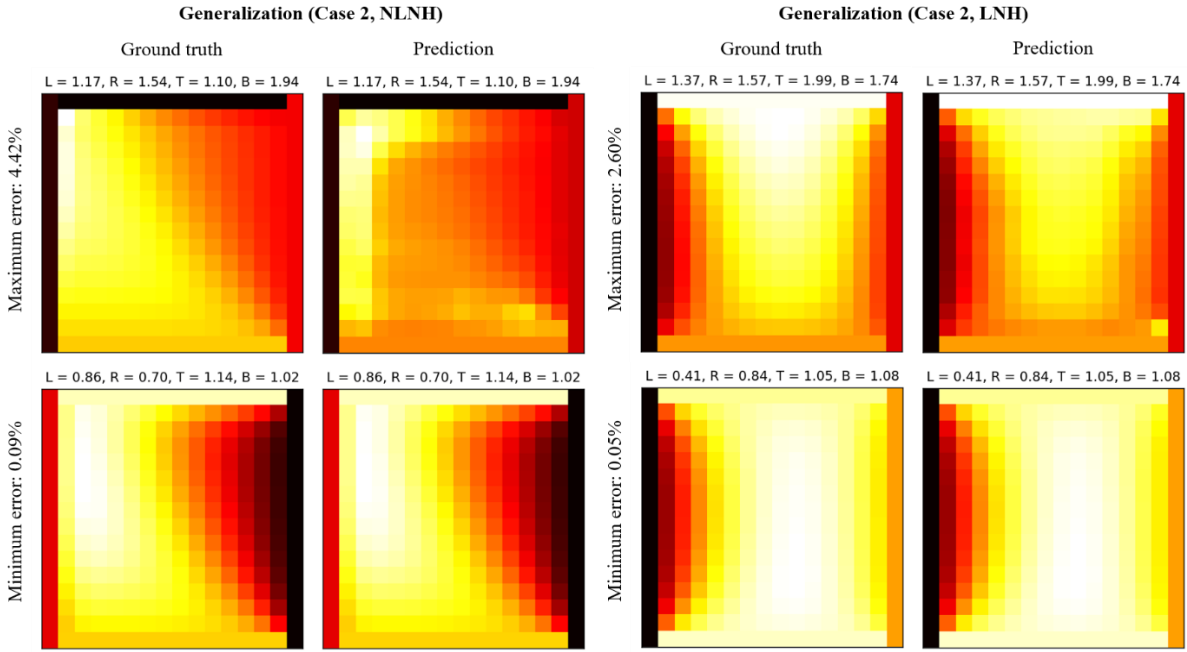

Figure 7. FDM results (ground truth) versus DNN output (prediction) for NLNH equation (left panel) and LNH equation (right panel): the upper and lower rows show the results for the maximum and minimum errors within the 300 input images whose randomly assigned BC values are in the range of [0.4, 1.4), [0.6, 1.6), [1, 2), and [1, 2), for the left (L), right (R), top (T), and bottom (B) boundaries, respectively.

**Conclusion**

Deep CNN networks based on the U-Net architecture with custom, physics-informed loss functions are implemented for three different PDEs. Inserting the physics of the problem into the networks, enabled us to work without labeled data. The only necessary input to train the DNNs are solution domains with various sets of BCs in the form of images that can be generated on the fly. The precision of the networks' predictions is all above 99.9% for the train and test BCs, even for nonlinear and non-homogeneous PDEs. The precision is more than 98% when the BCs are generalized to the values that are not present during the training. These results show the power of integrating physics laws into the structure of the networks when trying to adopt ML for physics problems.

We note that BCs containing derivative terms (Neumann BCs) can be implemented with slight modifications to the network, in addition to Dirichlet BCs of the present study. Such BCs can be discretized and defined as the difference between the two outermost pixels of the input images for instance. The method described in this paper can be extended for a set of coupled PDEs by developing networks with multiple inputs and outputs. Although the governing PDEs were assumed to be known in the examples of this study, algorithms can be implemented to verify the existence of specific differential operators from a range of built-in ones, in a problem whose governing equation is unknown[10]. In this case, some labeled data is needed to figure out the form of the PDE before using the current approach to solve it.

In summary, we presented a simple DL model to solve challenging PDEs based on CNNs. Use of CNNs instead of networks with fully connected layers has the advantage of faster computation times and, therefore, is well aligned with the needs of target communities.